\documentclass[prl,twocolumn,showpacs,superscriptaddress,preprintnumbers,amssymb]{revtex4}
\usepackage{graphicx}
\usepackage{dcolumn}
\usepackage{bm}
\usepackage{wasysym}

\newcommand{\beq}{\begin{equation}}
\newcommand{\eeq}{\end{equation}}
\newcommand{\beqn}{\begin{eqnarray}}
\newcommand{\eeqn}{\end{eqnarray}}

\newcommand{\sgn}{{\rm sgn}}

\begin{document}
\title{Edge states generated by spin-orbit coupling at domain walls in magnetic semiconductors}
\author{Cenke Xu}
\affiliation{Department of Physics, University of California,
Berkeley, CA 94720}\affiliation{Kavli Institute of Theoretical
Physics, University of California, Santa Barbara, CA, 93106}
\author{J.~E.~Moore}
\affiliation{Department of Physics, University of California,
Berkeley, CA 94720} \affiliation{Materials Sciences Division,
Lawrence Berkeley National Laboratory, Berkeley, CA 94720}
\date{\today}
\begin{abstract}

Electronic states localized at domain walls between
ferromagnetically ordered phases in two-dimensional electron
systems are generated by moderate spin-orbit coupling. The spin
carried by these states depends on the slope of the magnetic
background at the domain wall.  The number of localized states is
determined by a real space topological number, and spin
perpendicular to the ferromagnetic order accumulates in these
localized states at domain walls.  These trapped states may be
observed in experiments that probe either spin density or
conduction paths in quantum
wells.

\end{abstract}
 \maketitle

Spin transport and spin dynamics in semiconductors have attracted
considerable recent interest.  Spin currents in response to an
applied electric field have been found in both hole-doped
semiconductors \cite{zhang2003} and the electron-doped
two-dimensional quantum well with structure inversion asymmetry in
the third direction \cite{niu2004}. These spin currents are
referred to as intrinsic spin currents in the sense that they do
not depend on relaxation by impurity scattering, and the resulting
``spin Hall effect'' preserves time-reversal symmetry, unlike the
ordinary Hall effect. However, in semiconductors such as GaAs
time-reversal symmetry can be easily broken if the sample is doped
with dilute Mn ions, since the Mn ions develop ferromagnetic order
even at rather low doping \cite{ohno1992, ohno1998}. The
ferromagnetic order couples to itinerant electrons or holes
through both the resulting magnetic field and local ``Hund's
rule'' coupling, and hence affects the motion of charge carriers;
an example is the anomalous Hall effect measured in experiments on
(Ga,Mn)As \cite{ohno1999}.

This work studies how magnetic domain formation modifies
conduction electron states in magnetic semiconductors that also
have spin-orbit coupling (recall that the existence of spin-orbit
coupling alone does not break time-reversal). In a sample with
spin-orbit coupling, the conduction electron magnetization near a
domain wall (the boundary between oppositely directed domains) has
a component perpendicular to the external field or the local
ferromagnetic order, due to one-dimensional electronic edge states
localized at the domain wall. The pair of chiral, spin-polarized
electron modes at an edge can also lead to an unconventional
spin-transport mechanism.  The mathematical origin of these domain
wall states is very similar to the origin of trapped edge states
in p-wave superconductors~\cite{monien} and their fractional
quantum Hall analogue~\cite{read2000}.

One possible way to detect the localized domain wall states we
find is through spatial imaging of spin accumulation using Kerr
rotation microscopy \cite{sih2005}\cite{stephens2003}. Spin
accumulation changes the direction of polarization of incident
light. If the incident light is polarized along $\hat{z}$
direction, the bulk spin will not rotate the polarization; only
the domain wall states can induce a nonzero Kerr angle $\theta_K$.
By measuring the change of polarization of reflected light, the
accumulation of perpendicular spin can be detected at the domain
wall. Other possible ways to measure the domain wall states are
through spin-polarized tunneling or imaging of electronic
conduction paths~\cite{westervelt}.

Consider an n-doped narrow-gap semiconductor quantum well with an
asymmetric confining potential in the $\hat{z}$ direction (i.e.,
perpendicular to the well).  In this case, the conducting electron
band has a Rashba spin-orbit coupling \cite{Rashba1960} term \beq
H_{so} = \alpha(k_x\sigma^y - k_y\sigma^x)\label{rashbaso} \eeq In
typical GaAs wells, $\alpha$ is about $10^{-12}\ $ eV-m
\cite{rashba1984}\cite{lommer1988}, which is quite weak. However,
in some materials, for instance HgTe, the Rashba coupling $\alpha$
is about 100 times larger than in
GaAs\cite{culcer2003}\cite{radantsev2001}, and since the Rashba
spin-orbit coupling is linear instead of quadratic with momentum,
the spin-orbit coupling will become the dominant term in the
Hamiltonian as long as $k$ is small. Furthermore, if the sample is
doped with Mn ions, then time-reversal symmetry could be broken by
the long-ranged order of local moments of Mn ions. Mn ions are
acceptors in semiconductor, but the physics discussed here can be
realized if more donators are also doped other than Mn ions.
Assume for simplicity that the Mn ions develop ferromagnetic order
in the $\hat{z}$ direction; there will then be Hund's-rule
coupling between the local Mn ion and itinerant electrons, i.e.,
the local energy of an electron depends on whether its spin is
parallel or antiparallel to the local moment. We neglect until the
end of this paper orbital magnetic effects, as the exchange energy
scale that determines the Hund's rule coupling is frequently
dominant.  The full Hamiltonian to quadratic order in $k$  is then
given by \beq H = \frac{k^2}{2m^\ast} + \alpha(k_x\sigma^y -
k_y\sigma^x) + V(x)\sigma^z. \label{rashbahunds}\eeq

Let us assume for the moment that the Rashba coupling is big
enough to dominate the quadratic term in the Hamiltonian
(\ref{rashbahunds}); the legitimacy of doing this will be checked
later. The Hamiltonian now reads \beq H = \alpha(k_x\sigma^y -
k_y\sigma^x) + V(x)\sigma^z \label{Dirac}\eeq If $V$ is a
constant, this is exactly the Hamiltonian for massive Dirac
fermions with mass gap $V$. Suppose now $V(x)$ has a domain wall
(a change of sign) at $x = 0$. In order to solve the spectrum of
this Hamiltonian, each electron state is a spinor
with components $a(x)$ and $b(x)$: \beqn \psi(x,y) =
e^{ik_yy}\left(
\begin{array}
{cc} a(x)\\
b(x)\\
\end{array}
\right). \eeqn

\begin{figure}
\includegraphics[width=3.0in]{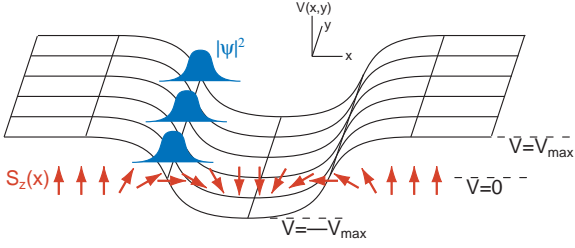}
\caption{ The spin of localized domain wall states
depends on the slope of the Zeeman field $V$.  In
this example, $V$ is a function of $x$ alone, and the
edge state is spin-polarized along either ${\hat x}$ or $-{\hat x}$.} \label{fig}
\end{figure}

Since $k_y$ is a conserved quantity, the Schr\"{o}dinger equation
is given by \beqn \alpha(-\partial_xb - k_yb) = (E - V(x))a\cr
\alpha (\partial_xa - k_ya) = (E + V(x))b \label{equation}\eeqn
This equation has two groups of solutions: \beqn a = - b \sim
e^{-\int^x_0\frac{V(x)}{\alpha} dx }, E = \alpha k_y\cr\cr a = b
\sim e^{\int^x_0\frac{V(x)}{\alpha} dx }, E = -\alpha k_y.
\label{solution}\eeqn If $V > 0$, which means $V(x)$ is changing
from negative to positive crossing the domain wall along
$+\hat{x}$, only the first solution is a localized state, and it
takes spin $\sigma^x = -1$, and the velocity of these states are
along $ + \hat{y}$. On the contrary, if $V < 0$, and $V(x)$
changes from positive to negative, the localized state is the
second solution above, with velocity along $-\hat{y}$ and spin
$\sigma^x = 1$. These eigenstates of the Hamiltonian indicates
that there is $\sigma^x$ accumulation at the domain wall, with the sign depending on the slope of the local ferromagnetic
order (Fig. \ref{fig}).

Note that (\ref{Dirac}) is just a Hamiltonian for massive Dirac
fermions with variable mass $V$. That the excitation gap closes at the
domain wall of $V$ is not so surprising, but the explicit
solution suggests that these massless domain wall states are
chiral, with spin depending on the direction of velocity of these
states.  However, it is important to note that periodicity of the band structure
generates an additional set of states with opposite chirality, as discussed below.
The situation here is analogous to the edge
states of $p\pm ip$ superconductors, in which the BdG equation breaks into two component equations
\cite{read2000}.



So far the $k^2$ term has been ignored in all the calculations.
When is this approximation legitimate? The term linear in $k$ will
always dominate quadratic terms when all the $k$ involved are
small. The characteristic $k$ involved is determined by the
localization length of this localized state, which is $\alpha/V$.
Hence the $k^2$ term is expected to be a small perturbation when
$(V/\alpha)^2/m$ is smaller than $\alpha\times V/\alpha$, if the
value of $\alpha$ in HgTe is taken
\cite{culcer2003}\cite{radantsev2001}, then $V$ should be smaller
than $10$ meV, which is a typical value for magnetic
semiconductors. The localization length of this state is of order
10 nm.

We now give a linear stability analysis to justify the above heuristic estimate.
We have also solved model band structures for small systems to verify that trapped
states indeed exist when predicted by the above intuitive criterion.
Suppose that the $k$ linear term numerically dominates the $k^2$
term; the stability of the localized solution for a domain wall
state can be checked by solving the Schr\"{o}dinger equation to
linear order in $1/m$ (this linear stability is in the same
spirit as the WKB method in quantum mechanics). For simplicity, $V(x)$
is taken as a step function $V(x) = V_0 \sgn(x)$ with $V_0
> 0$. $\sgn(x)$ is the sign function of $x$, $\sgn(x) = +1$ when
$x
> 0$, and $V(x) = -1$ when $x < 0$. We assume at first order in $1/m$
that the two components of the solution of the Schr\"{o}dinger equation
are modified to be \beqn \psi(x,y) = e^{k_yy}\left(
\begin{array}
{cc} (1 + f(x)/m)\exp(-\int dx\,V(x)/\alpha)\\
- (1 + g(x)/m)\exp(-\int dx\,V(x)/\alpha)
\\
\end{array}
\right) \label{stability}\eeqn and that the energy is slightly
changed to $E = \alpha k_y + \delta E$. As long as $f$ and $g$
have solutions that are polynomials of $x$, the localized states
are still stable. After solving to first order in $1/m$, $f$ and
$g$ have the following form: \beqn f(x) = \frac{V}{\alpha^2}(
k_y|x| + \frac{1}{2}\sgn(x)) + C \cr\cr g(x) = \frac{V}{\alpha^2}(
k_y|x| - \frac{1}{2}\sgn(x)) + C \eeqn These are solutions with $
\delta E = 1/m(V^2/\alpha^2 + k_y^2)$.  These solutions suggest
that the chiral localized states are stable even when there is a
numerically small $k^2$ term.

The chiral domain wall localized states derived so far are not the
only states localized at the domain wall, although they are the
most obvious ones. Had the original model been defined on the 2d square lattice, the
Rashba model would read \beqn H &=& -t (\cos k_x + \cos k_y -2) +
\alpha (\sin k_x\sigma^y - \sin k_y \sigma^x) \cr&&+ V\sigma^z.
\label{lahamiltonian}\eeqn
The existence of an additional set of states is familiar from lattice simulations
of chiral fermions: here, the result is an additional set of trapped states at a different momentum from the original states,
as now shown.

The lattice model coefficient $\alpha$ has different units than in
the continuum theory. If the continuum spin-orbit coupling is
$10^{-12}$ eV-m , and the lattice constant is $10^{-10}$m, the
lattice model coefficient $\alpha$ is about $10$meV. The localized
states in previous paragraphs were obtained from linearizing the
Hamiltonian around momentum $(0,0)$. However, the Hamiltonian
(\ref{lahamiltonian}) can always be expanded around an arbitrary
momentum $\vec{k}_{0x}$ by defining the slow wave function
$\tilde{\psi}(x)$ from $\psi(x) = e^{i k_{0x} x}\tilde{\psi}(x)$.

After linearizing at momentum $k_{0x}$, the Hamiltonian
(\ref{lahamiltonian}) becomes \beqn H &=& tk_x\sin k_{0x} -t\cos
k_y -t(\cos k_{0x} -2) + V\sigma^z \cr &&+\alpha(k_x\cos
k_{0x}\sigma^y - \sin k_y\sigma^x) + \alpha\sin
k_{0x}\sigma^y.\label{linearla}\eeqn Here $(k_x,k_y)$ is the
momentum of the slow mode $\tilde{\psi}$. If we ignore the last
term $\alpha\sin k_{0x}\sigma^y$ first (the legitimacy will be
discussed later), the resulting Schr\"{o}dinger equations for the
two-component spinor $\tilde{\psi} = (a, b)^T$ read \beqn -it\sin
k_{0x}\partial_x a - \alpha\cos k_{0x}\partial_x b - \alpha\sin
k_y b \cr = (E + t\cos k_{0x} + t\cos k_{y} - 2t -V)a, \cr\cr
-it\sin k_{0x}\partial_x b + \alpha\cos k_{0x}\partial_x a -
\alpha\sin k_y a \cr = (E + t\cos k_{0x} + t\cos k_{y} - 2t + V)b.
\eeqn

For simplicity, assume that $V > (<) 0$ when $x < (>) 0$ These two
equations have localized solutions as follows: \beqn E &=& -
\frac{\sqrt{\alpha^2\cos^2 k_{0x} - t^2\sin^2 k_{0x}}}{\cos
k_{0x}}\sin k_{y} \cr && - t\cos k_{0x} - t\cos k_{y} + 2t; \cr
\cr b &=& \gamma a \sim \exp(\int^x dx
\frac{V}{\sqrt{\alpha^2\cos^2 k_{0x} - t^2\sin^2 k_{0x}}}); \cr
\cr \cr \langle \sigma^x \rangle &=& \frac{2Re(\gamma)}{1 +
|\gamma|^2}, \langle \sigma^y \rangle = \frac{2Im(\gamma)}{1 +
|\gamma|^2}, \langle \sigma^z \rangle = \frac{1 - |\gamma|^2}{1 +
|\gamma|^2}; \cr \cr \cr \gamma &=& \frac{\sqrt{\alpha^2\cos^2
k_{0x} - t^2\sin^2 k_{0x}} - it\sin k_{0x}}{\alpha\cos
k_{0x}}.\label{solution2}\eeqn We see for the localized states
$\sin k_{0x}$ is about $\alpha / t$, which is generally very
small, therefore the last term in (\ref{linearla}) is a small
perturbation to the whole Hamiltonian. The stability of these
localized states when $k^2$ and last term in (\ref{linearla}) is
taken into account can be proved in the same WKB manner as we
derived formula (\ref{stability}).

As a check, we can assume $k_{0x} = 0$ and obtain the same answer
as (\ref{solution}). All these localized states tend to align spin
along $+\hat{x}$. These states are only localized when
$\sqrt{\alpha^2\cos^2 k_{0x} - t^2\sin^2 k_{0x}}$ is a real
number, which means \beq |\tan k_{0x}| < \alpha / t.
\label{condition}\eeq However, in order to make sure the
approximate solutions in (\ref{solution2}) are approximately
orthogonal with each other, $k_{0x}$ cannot be taken as continuous
value. The intervals between each $k_{0x}$ should be at least
$2\pi V/\alpha$, which is the inverse of localization length.
Therefore the total number of localized channels close to energy
$E_f = 0$ is $N \approx 1 + \alpha^2/(2\pi tV)$. There are more
than one localized channel if $\alpha^2/(tV) > 2\pi$, which can
possibly be satisfied in systems with large spin-orbit coupling,
like HgTe. States with $k_{0x}$ within the intervals $(-\alpha/t,
\alpha/t)$ and $(\pi - \alpha/t , \pi + \alpha / t)$ satisfy the
condition (\ref{condition}), but according to the energy spectrum
in (\ref{solution2}), the states close to $k_{0x} = \pi$ have
energy higher than states close to $(0,0)$ by $2t$, and the whole
spectrum width of these localized states is about $2\pi\alpha$. If
the Fermi energy is fixed at $E_f = 0$ and $t > \pi\alpha$, only
localized states close to $k_{0x} = 0$ should be relevant. These
states have the same spin alignment as in Figure. \ref{fig}, but,
because the energy $E$ is a periodic function on the Brillouin
zone, there is always another branch of states with velocity
opposite to the ones obtained in the continuum states. The
velocity is $v_y = \partial E / \partial k_y$, and since $E$ is a
periodic function of $k_y$, after integrating over the whole
Brillouin zone the spin current $1/2\{s^x, v_y\}$ vanishes. For
example, if $t = 0$, $E = - \alpha \sin k_y$, states close to $k_y
= \pi$ have opposite velocity from states close to $k_y = 0$.

Many types of edge states are related to topology in momentum
space \cite{volovik2005}.  Examples are the quantum Hall
effect\cite{wen1990}, and some quantum spin Hall models
\cite{kane2005,zhang2005,haldane2005,xum2005,wu2005}. The usual
way to bridge the bulk momentum-space topology and edge modes is
through Chern-Simons theory \cite{wen1991}, because Chern-Simons
theory is gauge invariant up to a boundary term, which gives rise
to chiral boson excitations at the boundary.  Momentum-space
topology is usually obtained in insulators, in which the whole
Brillouin zone is filled; since physical quantities have the
periodicity of the Brillouin zone, integrals over the whole
Brillouin zone are integrals on a compact manifold, which can give
rise to a nontrivial topological number.  Here the situation is
slightly different: there are localized domain wall modes, but the
number of edge states at a single domain wall is determined by the
number of bands with linear (Dirac) spectrum near the wall, which
results when a band is near a minimum or maximum of the band
structure neglecting spin-orbit coupling.  Note that the number of
domain walls is related to the number of zero crossings of the
Zeeman field: to count this explicitly, a vector $\vec{N}$ has to
be defined. Let the $\hat{z}$ component of $\vec{N}$ be $N_z(x) =
V(x)$, and the in-plane part of the vector be $\vec{N}_{in} =
\vec{\nabla}V$. The number of walls is the Skyrmion number \beq n
= \frac{1}{2\pi}\int
d^2x\hat{N}\cdot(\partial_x\hat{N}\times\partial_y\hat{N}).
\label{skyrmion}\eeq Here $\hat{N}$ is the unit vector along
$\vec{N}$. This counting breaks down when states from different
walls begin to mix. So the number of localized states is
determined by this real space topological number. The spin
polarization in Fig. \ref{fig} also has nonzero real space
Skyrmion number, which is equal to (\ref{skyrmion}).


The domain wall states affect various physical quantities. The
most direct result is the $\sigma^x$ magnetization. Also, once an
electric field is turned on, charge current at the domain wall
carries spin current (i.e., the charge current is spin-polarized),
with spin maximally polarized perpendicular to the ferromagnetic
order. If the width of the sample (or the length of the domain
wall) $L_y$ is shorter than the mean free length of the electrons,
the modes at $k_{0x} = 0$ will induce ballistic spin current
conductance similar to the weakly quantized noninteracting
electric conductance in 1d systems: for voltage $U$, $I_s =
\frac{1}{2}\frac{e}{2\pi} U$, and the total spin current
conductance is obtained from summing over all the $k_{0x}$ that
satisfy (\ref{condition}).  When $\alpha / t < 1$, the result is $
Ne/4\pi $, with $N$ the number of localized channels.

The existence of spin accumulation and spin-polarized currents can be
understood from symmetry arguments.  In a 2d plane with asymmetric
confining potential in $\hat{z}$ as well as uniform Zeeman field
$V \hat{z}$, there are the following symmetries: $\{TP_x, TP_y, P_xP_y
\}$. Here $P_x$ means $x
\rightarrow - x$, $P_y$ means $y \rightarrow -y $, and $T$ means time-reversal.
Because $T$, $P_x$ and $P_y$ all flip $V \hat{z}$, any
combination of two of them should be a symmetry
transformation of the system.

Therefore, spin polarization $\langle s^x \rangle$ and $\langle
s^y \rangle$ are not allowed by these three symmetries. Instead,
spin currents $j^x_y = 1/2\{s^x,v_y\}$ and $j^y_x =
1/2\{s^y,v_x\}$ are allowed by symmetry. However, at the domain
wall along $\hat{y}$, if we assume $V(x)$ is an odd function, the
symmetries are $\{ P_x, TP_y\}$. Now both $\langle s^x \rangle$
and $j^x_y$ are allowed by symmetry, consistent with the
microscopic calculation given above.

For comparison to experiment, let us estimate the relative
contribution to the electronic density of states from the
localized states. For an estimate, suppose that the Rashba
spin-orbit coupling in continuum theory is $10^{-10}$ eV-m (as in
HgTe), the effective mass is taken as $0.1\ m_e$, and if $V$ is
about $10$meV, there can be more than one independent localized
channels at every domain wall. Then the spin density $\langle
\sigma^x \rangle$ per length at the wall is about $10^{8}\ \hbar$
cm$^{-1}$ times the number of channels.

We also find perpendicular spin polarization at domain walls when
time-reversal symmetry of the itinerant electrons is broken by an
orbital magnetic field rather than by a Zeeman field. The
Hamiltonian including spin-orbit coupling for a 2D electron gas in
magnetic field is \beqn H = \alpha((k_x - eA_x)\sigma^y - (k_y -
eA_y)\sigma^x) + B(x) g \mu_B s^z.\label{mag} \eeqn The domain
wall can be treated as the edge of two quantum Hall systems, and
as is well known, the electron states at the edge are already
chiral in the absence of spin-orbit coupling. The spin direction
of these domain wall states can be guessed by taking the square of
(\ref{mag}): any eigenstate of $H$ is an eigenstate of $H^2$,
although the converse is false.  (We also ignored the quadratic
term in the Hamiltonian, since the spin-orbit part will be enough
for an intuitive discussion.) The square of (\ref{mag}) is \beq
H^2 = \alpha^2(k_x^2 + (k_y - eA_y)^2) + 2eB\alpha^2\sigma^z +
2\alpha {\partial_x B} \mu_Bs^x, \eeq where the Landau gauge $A_x
= 0, A_y = \int^x B dx$ has been taken and $g=2$. Far away from
the domain wall, where the magnetic field is varying slowly, $A_y$
can be taken as $Bx$. The first two terms are familiar to all,
they are just the nonrelativistic particles moving in magnetic
field. The third term results from the slope of the magnetic
field, and pushes the spin along $x$, perpendicular to the
magnetic field. Detailed calculation of the full Hamiltonian gives
qualitatively the same result: spin-orbit coupling leads to
confined, maximally spin-polarized edge states in the Zeeman case
(which is expected to be more relevant to experiment), and also
tends to spin-polarize ordinary quantum Hall edge states in the
orbital case, perpendicular to the domain wall.

The authors wish to acknowledge conversations with D.-H. Lee and support from NSF grant DMR-0238760.

\bibliography{domain}
\end{document}